\documentclass[twocolumn,showpacs,preprintnumbers,amsmath,amssymb,prb]{revtex4}

\usepackage{graphicx}% Include figure files
\usepackage{dcolumn}% Align table columns on decimal point
\usepackage{bm}% bold math
\usepackage{tabularx}% Table width
\usepackage{amsmath}% Higher math

\begin{document}

\title{Evidence for nodeless superconducting gap in NaFe$_{1-x}$Co$_x$As from low-temperature thermal conductivity measurements}

\author{S. Y. Zhou,$^1$ X. C. Hong,$^1$ X. Qiu,$^1$ B. Y. Pan,$^1$ Z. Zhang,$^1$ X. L. Li,$^1$ W. N. Dong,$^1$ \\ A. F. Wang,$^2$ X. G. Luo,$^2$ X. H. Chen,$^2$ and S. Y. Li$^{1,*}$}

\affiliation{$^1$State Key Laboratory of Surface Physics, Department
of Physics, and Laboratory of Advanced Materials, Fudan University,
Shanghai 200433, P. R. China\\
$^2$Hefei National Laboratory for Physical Science at Microscale and
Department of Physics, University of Science and Technology of
China, Hefei, Anhui 230026, China}

\date{\today}

\begin{abstract}
The thermal conductivity of optimally doped
NaFe$_{0.972}$Co$_{0.028}$As ($T_c \sim$ 20 K) and overdoped
NaFe$_{0.925}$Co$_{0.075}$As ($T_c \sim$ 11 K) single crystals were
measured down to 50 mK. No residual linear term $\kappa_0/T$ is
found in zero magnetic field for both compounds, which is an
evidence for nodeless superconducting gap. Applying field up to $H$
= 9 T ($\approx H_{c2}/4$) does not noticeably increase $\kappa_0/T$
in NaFe$_{1.972}$Co$_{0.028}$As, which is consistent with multiple
isotropic gaps with similar magnitudes. The $\kappa_0/T$ of
overdoped NaFe$_{1.925}$Co$_{0.075}$As shows a relatively faster
field dependence, indicating the increase of the ratio between the
magnitudes of different gaps, or the enhancement of gap anisotropy
upon increasing doping.

\end{abstract}


\maketitle

The discovery of high-temperature superconductivity in iron
pnictides \cite{Kamihara,XHChen} has attracted great attentions.
Many efforts have been devoted to determine the symmetry and
structure of their superconducting gap, \cite{Hirschfeld} which is
one of the keys to understand the electronic pairing mechanism.
\cite{FaWang}

For the most studied (Ba,Sr,Ca,Eu)Fe$_2$As$_2$ (122) system, while
multiple nodeless superconducting gaps were demonstrated near
optimal doping, \cite{HDing,KTerashima,XGLuo,LDing,Tanatar1} nodal
superconductivity was found in the extremely hole-doped
KFe$_2$As$_2$, \cite{JKDong1,KHashimoto1} and highly anisotropic gap
\cite{Tanatar1} or isotropic gaps with significantly different
magnitudes \cite{JKDong2,YBang} were suggested in the heavily
electron-doped Ba(Fe$_{1-x}$Co$_x$)$_2$As$_2$. Intriguingly, nodal
superconductivity was also found in isovalently doped
BaFe$_2$(As$_{1-x}$P$_x$)$_2$ and Ba(Fe$_{1-x}$Ru$_x$)$_2$As$_2$.
\cite{YNakai,KHashimoto2,YZhang,XQiu}

For the two stoichiometric LiFeAs and LiFeP (111) compounds, it was
clearly shown that the former has nodeless superconducting gaps,
\cite{Borisenko,Tanatar2,HKim,KUmezawa} and the latter has nodal
gap. \cite{KHashimoto3} However, for another 111 compound
NaFe$_{1-x}$Co$_x$As, recently there were controversial experimental
results on its gap structure. \cite{ZHLiu,KCho} High-resolution
angle-resolved photoemission spectroscopy (ARPES) measurements on
NaFe$_{0.95}$Co$_{0.05}$As ($T_c \sim$ 18 K) unambiguously showed
nearly isotropic superconducting gaps on all three Fermi surfaces,
\cite{ZHLiu} but London penetration depth measurements claimed nodal
gap in NaFe$_{1-x}$Co$_x$As from underdoped to overdoped range ($x$
ranging from 0.02 to 0.10). \cite{KCho} To resolve this important
issue, more experiments are highly desired.

The ultra-low-temperature heat transport measurement is a bulk
technique to probe the gap structure of superconductors.
\cite{Shakeripour} Whether there is a finite residual linear term
$\kappa_0/T$ in zero field is a good judgement on the existence of
gap nodes or not. The field dependence of $\kappa_0/T$ can further
give information of nodal gap, gap anisotropy, or multiple gaps.
\cite{Shakeripour} During past three years, thermal conductivity
measurements have provided valuable experimental results to clarify
the gap structure of iron-based superconductors. \cite{Stewart}

In this Rapid Communication, we present the thermal conductivity
measurements of optimally doped NaFe$_{0.972}$Co$_{0.028}$As and
overdoped NaFe$_{0.925}$Co$_{0.075}$As single crystals down to 50
mK. We find no evidences for gap nodes in both compounds, which
supports previous ARPES results in Ref. 23, and disagrees with the
penetration depth experiments in Ref. 24. Furthermore, the
relatively fast field dependence of $\kappa_0/T$ in
NaFe$_{0.925}$Co$_{0.075}$As indicates the increase of the ratio
between the magnitudes of those isotropic gaps, or the enhancement
of gap anisotropy upon increasing doping. This gap structure
evolution is similar to that of Ba(Fe$_{1-x}$Co$_x$)$_2$As$_2$
system.

The NaFe$_{1-x}$Co$_x$As single crystals were synthesized by flux
method with NaAs as the flux. \cite{CHe} NaAs was prepared by
reacting the Na chunks and As powders in the evacuated quartz tube
at 200 $^o$C for 10 hours. Then the powders of NaAs, Fe and Co were
mixed together according to the ratio NaAs:Fe:Co = 4:1-$x$:$x$. The
mixture was placed into an alumina crucible and then sealed inside
an iron crucible under Ar atmosphere. The crucible was put in Ar
filled tube furnace and slowly heated up to 950 $^o$C and kept for
10 hours, then slowly cooled down to 600 $^o$C at a rate of 3
$^o$C/h. The shining plate-like NaFe$_{1-x}$Co$_x$As single crystals
with maximum size of 5$\times$5$\times$0.2 mm$^3$ were obtained. The
actual chemical composition of the single crystals is determined by
energy-dispersive X-ray spectroscopy. The cobalt contents are $x$ =
0.028 and 0.075, respectively, for the samples with nominal $x$ =
0.05 and 0.20. The dc magnetic susceptibility was measured at $H$ =
20 Oe, with zero-field cooled, using a SQUID (MPMS, Quantum Design).

The single crystals were stored and shipped under the environment of
inert gas. After exposed in air, the sample was quickly cleaved to a
rectangular shape with typical dimensions of $\sim$2.50$\times$1.00
mm$^2$ in the $ab$-plane, and $\sim$100 $\mu$m along the $c$-axis.
Then four silver wires were attached on the sample with silver
paint, which were used for both thermal conductivity and resistivity
measurements. The time of exposure in air is less than one hour,
before the heat transport study. In-plane thermal conductivity was
measured in a dilution fridge using a standard four-wire
steady-state method with two RuO$_2$ chip thermometers, calibrated
{\it in situ} against a reference RuO$_2$ thermometer. After the
thermal conductivity measurements, the sample was quickly (roughly
30 minutes exposure in air) switched to a $^4$He cryostat for
resistivity measurements. Finally, the contact resistances were
examined, with typical value of 100 m$\Omega$ at 2 K. Magnetic
fields were applied along the $c$-axis. To ensure a homogeneous
field distribution in the samples, all fields were applied at
temperature above $T_c$.

\begin{figure}
\includegraphics[clip,width=5.5cm]{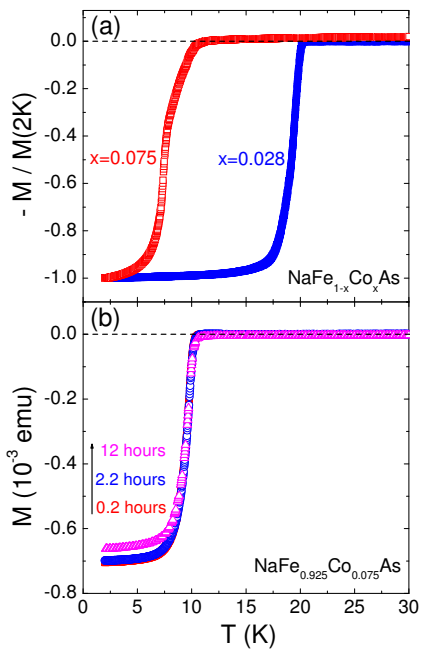}
\caption{(Color online). (a) Normalized dc magnetization of
NaFe$_{0.972}$Co$_{0.028}$As and NaFe$_{0.925}$Co$_{0.075}$As single
crystals used for transport study. (b) Magnetization of another
NaFe$_{0.925}$Co$_{0.075}$As single crystal from the same batch,
which also has millimeter size in the $ab$-plane and $\sim$100
$\mu$m along $c$-axis, after being exposed in air for 0.2, 2.2, and
12 hours, respectively. The magnetization barely changes during the
initial 2.2 hours.}
\end{figure}

\begin{figure}
\includegraphics[clip,width=5.08cm]{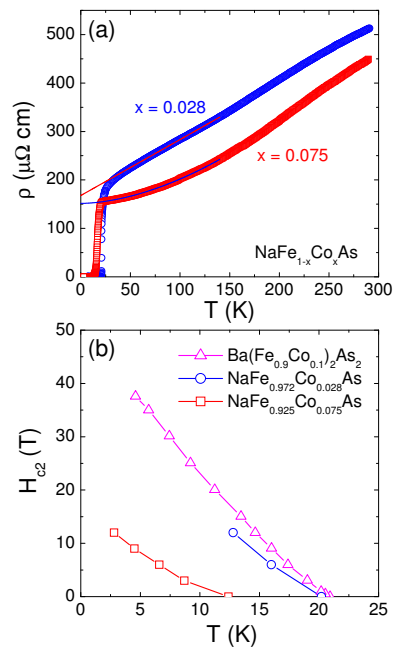}
\caption{(Color online). (a) In-plane resistivity of
NaFe$_{0.972}$Co$_{0.028}$As and NaFe$_{0.925}$Co$_{0.075}$As single
crystals. The solid lines are linear fit for
NaFe$_{0.972}$Co$_{0.028}$As between 50 and 140 K, and a fit to
$\rho(T) = \rho_0 +AT^n$ for NaFe$_{0.925}$Co$_{0.075}$As between 25
and 140 K, respectively. (b) The upper critical field $H_{c2}(T)$ of
NaFe$_{0.972}$Co$_{0.028}$As and NaFe$_{0.925}$Co$_{0.075}$As,
defined by $\rho = 0$. For comparison, the data of
Ba(Fe$_{0.9}$Co$_{0.1}$)$_2$As$_2$ with $T_c \approx 22$ K is also
plotted. \cite{AYamamoto} }
\end{figure}

Figure 1(a) presents the normalized dc magnetization of
NaFe$_{0.972}$Co$_{0.028}$As and NaFe$_{0.925}$Co$_{0.075}$As single
crystals. The onset transition temperatures $T_c$ are 20 and 11 K,
respectively. For NaFe$_{0.972}$Co$_{0.028}$As, the data were taken
on a crystal right after being cleaved from the sample used for
transport study. For NaFe$_{0.925}$Co$_{0.075}$As, the data were
taken directly on the sample used for transport study, after all
transport experiments have been done, totally about 2 hours exposure
in air. In order to check how sensitive is the bulk
superconductivity to the exposure time, we measured another
NaFe$_{0.925}$Co$_{0.075}$As single crystal from the same batch,
which also has millimeter size in $ab$-plane and $\sim$ 100 $\mu$m
along $c$-axis. Fig. 1(b) shows its magnetization after being
exposed in air for 0.2, 2.2, and 12 hours, respectively. The
magnetization barely changes during the initial 2.2 hours, therefore
it is concluded that within 2.2 hours there is little effect of air
on the bulk superconductivity of NaFe$_{1-x}$Co$_x$As single
crystals with this kind of size. We notice that the bulk
superconductivity degrades faster in air for smaller and thinner
single crystals, which is understandable. Since the contact
resistances remain low at the end of transport measurements, our
resistivity and thermal conductivity data represent the intrinsic
properties of the two NaFe$_{1-x}$Co$_x$As compounds.

Figure 2(a) shows the in-plane resistivity $\rho(T)$ of
NaFe$_{0.972}$Co$_{0.028}$As and NaFe$_{0.925}$Co$_{0.075}$As single
crystals. The $T_c$ defined by $\rho = 0$ are 20.2 and 12.4 K,
respectively. This $T_c$ of NaFe$_{0.925}$Co$_{0.075}$As sample is
about 1 K higher than that determined from magnetization
measurement, possibly due to slight inhomogeneity. We name these two
samples as OP20K and OD11K. For OP20K, the resistivity data between
50 and 140 K can be fitted linearly, giving a residual resistivity
$\rho_0$ = 167.8 $\pm$ 0.1 $\mu\Omega$cm. For OD11K, the data
between 25 and 140 K are fitted to $\rho(T) = \rho_0 +AT^n$, which
gives $\rho_0$ = 151.6 $\pm$ 0.1 $\mu\Omega$cm and $n$ = 1.76 $\pm$
0.01. The non-Fermi-liquid linear behavior of $\rho(T)$ near optimal
doping, and the increase of power $n$ in the overdoped regime have
been observed in BaFe$_2$(As$_{1-x}$P$_x$)$_2$, which was considered
as the signature of a quantum critical point. \cite{SKasahara}

\begin{figure}
\includegraphics[clip,width=5.5cm]{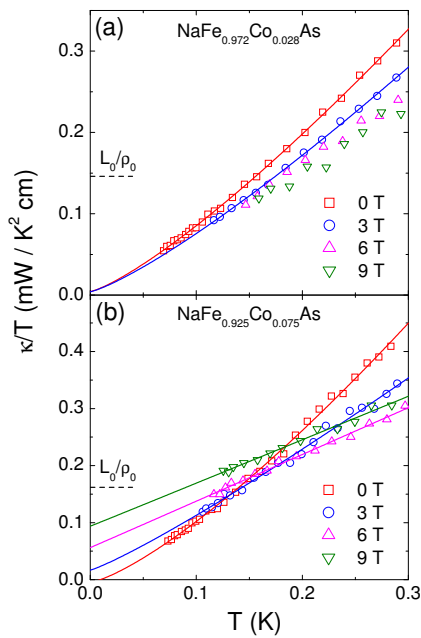}
\caption{(Color online). Low-temperature in-plane thermal
conductivity of NaFe$_{0.972}$Co$_{0.028}$As and
NaFe$_{0.925}$Co$_{0.075}$As in zero and magnetic fields applied
along the $c$-axis. The solid lines are fits to $\kappa/T = a +
bT^{\alpha-1}$. The dash lines are the normal-state Wiedemann-Franz
law expectation $L_0$/$\rho_0$, with $L_0$ the Lorenz number 2.45
$\times$ 10$^{-8}$ W$\Omega$K$^{-2}$ and $\rho_0$ = 167.8 and 151.6
$\mu\Omega$cm, respectively.}
\end{figure}

\begin{figure}
\includegraphics[clip,width=6.5cm]{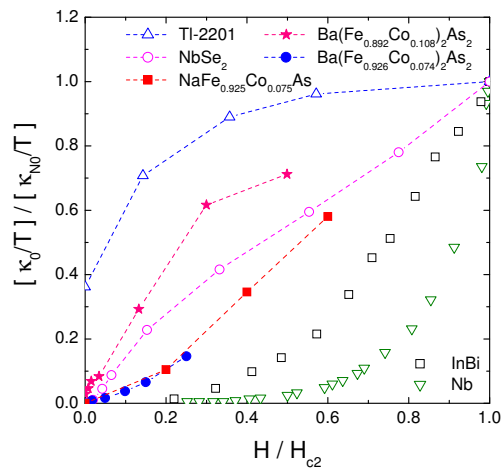}
\caption{(Color online). Normalized residual linear term
$\kappa_0/T$ of NaFe$_{0.925}$Co$_{0.075}$As as a function of
$H/H_{c2}$. For comparison, similar data are shown for the clean
$s$-wave superconductor Nb, \cite{Lowell} the dirty $s$-wave
superconducting alloy InBi, \cite{Willis} the multi-band $s$-wave
superconductor NbSe$_2$, \cite{Boaknin} an overdoped $d$-wave
cuprate superconductor Tl-2201, \cite{Proust} the optimally doped
Ba(Fe$_{0.926}$Co$_{0.074}$)$_2$As$_2$ and overdoped
Ba(Fe$_{0.892}$Co$_{0.108}$)$_2$As$_2$. \cite{Tanatar1}}
\end{figure}

The resistivity of these two samples were also measured in magnetic
fields up to $H$ = 12 T, in order to estimate the upper critical
field $H_{c2}$. Fig. 2(b) plots the temperature dependence of
$H_{c2}$ for OP20K and OD11K, defined by $\rho = 0$. For comparison,
the data of Ba(Fe$_{0.9}$Co$_{0.1}$)$_2$As$_2$ with $T_c \approx 22$
K is plotted. \cite{AYamamoto} From Fig. 2(b), we roughly estimate
$H_{c2} \approx$ 36 and 15 T for OP20K and OD11K, respectively. To
choose a slightly different $H_{c2}$ does not affect our discussion
on the field dependence of $\kappa_0/T$ below.

Figure 3 shows the temperature dependence of in-plane thermal
conductivity for OP20K and OD11K in zero and magnetic fields,
plotted as $\kappa/T$ vs $T$. To get the residual linear term
$\kappa_0/T$, we fit the curves to $\kappa/T$ = $a + bT^{\alpha-1}$,
in which the two terms $aT$ and $bT^{\alpha}$ represent
contributions from electrons and phonons, respectively.
\cite{Sutherland,SYLi} The power $\alpha$ of the phonon term is
typically between 2 and 3. \cite{Sutherland,SYLi} For OP20K in zero
field, the fitting gives $\kappa_0/T$ = $a$ = 0.004 $\pm$ 0.003 mW
K$^{-2}$ cm$^{-1}$, with $\alpha$ = 2.26 $\pm$ 0.03. For OD11K in
zero field, $\kappa_0/T$ = -0.005 $\pm$ 0.008 mW K$^{-2}$ cm$^{-1}$
and $\alpha$ = 2.32 $\pm$ 0.06 are obtained. Previously $\alpha
\approx$ 2.2 has been observed in BaFe$_2$As$_2$, \cite{Kurita} and
$\alpha \approx$ 2 was found in BaFe$_{1.9}$Ni$_{0.1}$As$_2$ and
KFe$_2$As$_2$. \cite{LDing,JKDong1}

Both $\kappa_0/T$ of OP20K and OD11K in zero field are negligible,
within our experimental error bar $\pm$ 0.005 mW K$^{-2}$ cm$^{-1}$.
\cite{SYLi} In nodelss superconductors, all electrons become Cooper
pairs as $T \to 0$, therefore there are no fermionic quasiparticles
to conduct heat and zero $\kappa_0/T$ is observed, as in the
conventional $s$-wave superconductor V$_3$Si. \cite{Sutherland} On
the contrary, in a nodal superconductor, the nodal quasiparticles
will contribute a finite $\kappa_0/T$ in zero field, as in the
$d$-wave cuprate superconductor Tl$_2$Ba$_2$CuO$_{6+\delta}$
(Tl-2201). \cite{Proust} Therefore the negligible $\kappa_0/T$ in
both OP20K and OD11K strongly suggest that their superconducting
gaps are nodeless. These results are consistent with previous ARPES
results in Ref. 23, but disagree with the London penetration depth
experiments in Ref. 24.

To gain further support for nodeless gap in OP20K and OD11K, we
check the field dependence of their $\kappa_0/T$. For OP20K, the fit
to the data in $H$ = 3 T also gives a negligible $\kappa_0/T$ =
0.004 $\pm$ 0.011 mW K$^{-2}$ cm$^{-1}$. Further increasing field,
the data in $H$ = 6 and 9 T become noisy and the fits are not very
good, but the tendency is that the $\kappa_0/T$ does not increase
noticeably up to $H$ = 9 T ($\approx H_{c2}/4$). This field effect
on $\kappa_0/T$ of OP20K is again consistent with the ARPES
experiments on optimally doped NaFe$_{1-x}$Co$_x$As, which show
isotropic gaps with $\Delta_1$ = 6.8 meV for the $\alpha$ hole band
and $\Delta_2$ = 6.5 meV for the $\gamma(\delta)$ electron bands.
\cite{ZHLiu} Since the magnitudes of these gaps are very close
($\Delta_1$/$\Delta_2$ = 1.05), the field effect on $\kappa_0/T$ is
just like that of a superconductor with single isotropic gap,
\cite{YBang} for examples, the clean $s$-wave superconductor Nb and
the dirty $s$-wave superconducting alloy InBi, shown in Fig. 4.
\cite{Lowell,Willis}

As seen in Fig. 3(b), the fits to the thermal conductivity data of
OD11K in fields look reliable, so all $\kappa_0/T$ are obtained up
to $H$ = 9 T. In Fig. 4, the normalized
$(\kappa_0/T)/(\kappa_{N0}/T)$ of OD11K is plotted as a function of
$H/H_{c2}$, with the normal-state Wiedemann-Franz law expectation
$\kappa_{N0}/T$ = $L_0$/$\rho_0$ = 0.162 mW K$^{-2}$ cm$^{-1}$ and
$H_{c2}$ = 15 T. Similar data of the multi-band $s$-wave
superconductor NbSe$_2$, \cite{Boaknin} Tl-2201, \cite{Proust} the
optimally doped Ba(Fe$_{0.926}$Co$_{0.074}$)$_2$As$_2$ ($T_c$ = 22.2
K) and overdoped Ba(Fe$_{0.892}$Co$_{0.108}$)$_2$As$_2$ ($T_c$ =
14.6 K) \cite{Tanatar1} are also plotted for comparison. Clearly,
the $\kappa_0(H)/T$ behavior of OD11K is very similar to that of
optimally doped Ba(Fe$_{0.926}$Co$_{0.074}$)$_2$As$_2$, and much
slower than that of overdoped
Ba(Fe$_{0.892}$Co$_{0.108}$)$_2$As$_2$.

For optimally doped BaFe$_{1.85}$Co$_{0.15}$As$_2$, the gaps are
isotropic, with $\Delta_1$ = 6.0 meV for the $\beta$ hole pocket and
$\Delta_2$ = 5.0 meV for the $\gamma(\delta)$ electron pockets.
\cite{KTerashima} The different magnitudes of those gaps
($\Delta_1$/$\Delta_2$ = 1.32) may result in the relatively fast
field dependence of $\kappa_0/T$ shown in Fig. 4. For overdoped
Ba(Fe$_{0.892}$Co$_{0.108}$)$_2$As$_2$, the $\kappa_0/T$ increases
very rapidly even at low field, which has been interpreted as the
manifestation of large ratio ($>$ 3) between $\Delta_1$ and
$\Delta_2$, \cite{JKDong2} or highly anisotropic gap.
\cite{Tanatar1} The similar $\kappa_0(H)/T$ behaviors between OD11K
and optimally doped Ba(Fe$_{0.926}$Co$_{0.074}$)$_2$As$_2$ suggest
that the ratio $\Delta_1$/$\Delta_2$ has increased from 1.05 in
OP20K to $\sim 1.3$ in OD11K, or some gaps become anisotropic up to
30\%. This needs to be checked by the ARPES experiments. From our
data, the gap structure evolution upon increasing doping is similar
in NaFe$_{1-x}$Co$_x$As and Ba(Fe$_{1-x}$Co$_x$)$_2$As$_2$ systems.
Note that the field dependence of $\kappa_0/T$ for OD11K again
disagrees with the claim of nodal gap in overdoped
NaFe$_{1-x}$Co$_x$As, \cite{KCho} since $\kappa_0(H)/T$ should
display a much faster field dependence in that case.

In summary, we have measured the thermal conductivity of optimally
doped NaFe$_{0.972}$Co$_{0.028}$As and overdoped
NaFe$_{0.925}$Co$_{0.075}$As single crystals down to 50 mK. The
absence of $\kappa_0/T$ in zero field for both compounds gives
strong evidence for nodeless gap. The field effect on $\kappa_0/T$
of NaFe$_{0.972}$Co$_{0.028}$As is consistent with multiple
isotropic gaps with similar magnitudes, as demonstrated by ARPES
experiments. The relatively faster field dependence of $\kappa_0/T$
for overdoped NaFe$_{0.925}$Co$_{0.075}$As suggests that the
different between the magnitudes of those gaps increases, or some
gaps become anisotropic upon increasing doping. Our results clearly
disagree with the claim of nodal gap in NaFe$_{1-x}$Co$_x$As based
on London penetration depth experiments, therefore the issue of gap
structure in this 111 system has been clarified.

This work is supported by the Natural Science Foundation of China,
the Ministry of Science and Technology of China (National Basic
Research Program No: 2009CB929203 and 2012CB821402), and the Program
for Professor of Special Appointment (Eastern Scholar) at Shanghai
Institutions of Higher Learning. \\

$^*$ E-mail: shiyan$\_$li@fudan.edu.cn

\end{document}